\documentclass[granma,referee]{svjour}
\usepackage{amssymb}
\usepackage{amsmath}
\usepackage{graphicx}
\usepackage{longtable}

\begin{document}

\title{Slow Steady-Shear of Plastic Bead Rafts}

\author{Michael Twardos and Michael Dennin}

\institute{Department of Physics and Astronomy, University of
California, Irvine. Irvine, CA 92696-4575}

\date{\today}

\maketitle

\begin{abstract}

Experimental measurements of the response of a two dimensional
system of plastic beads subjected to steady shear are reported.
The beads float at the surface of a fluid substrate and are
subjected to a slow, steady-shear in a Couette geometry.  The flow
consists of irregular intervals of solid-like, jammed behavior,
followed by stress relaxations. We report on statistics that
characterize the stress fluctuations as a function of several
parameters including shear-rate and packing density. Over a range
of densities between the onset of flow to the onset of buckling
(overpacking) of the system, the probability distribution for
stress fluctuations is essentially independent of the packing
density, particle dispersity, and interaction potential (varied by
changing the substrate). Finally, we compare the observed stress
fluctuations with those observed in other complex fluids.

\end{abstract}

\section{Introduction}


For equilibrium systems, it is well understood how fluctuations
are controlled by temperature. Especially important is the
understanding of the role of fluctuations in the modern theory of
phase transitions \cite{OA02}. Similarly, one expects
nonequilibrium fluctuations to play an equally important role in
transitions in driven systems, such as the proposed jamming
transition as a function of external stress
\cite{LN98,TPCSW01,ITPJamming}. In driven systems, the
fluctuations of interest are typically not thermal in nature.
Instead, they are the result of the external driving, and this
raises important questions and presents a number of challenges. As
a starting point, it is useful to characterize the statistics of
these ``athermal'' fluctuations and ask how similar such
statistics are to equivalent measures of thermal fluctuations.
Then, one can consider the impact athermal fluctuations on the
system's dynamics.

A general feature of fluctuations in many driven systems is the
occurrence of non-Gaussian features, typically an asymmetric shape
with exponential tails, in probability distributions of the
fluctuating quantity. It is interesting to ask under what
conditions such distributions are {\it universal}. This question
has been studied in the context of turbulent fluid flows for
thermally driven fluids \cite{CL98} (thermal convection) and
electrically driven fluids \cite{GGK01,TG03} (electroconvection).
More recent work has considered this question in shaken granular
systems \cite{FM03}. The focus in these studies has been on
fluctuations in the dissipated power. Two different approaches for
characterizing fluctuations in the dissipated power were used. The
first is based on the generalized fluctuation-dissipation theorem
of Gallavotti and Cohen, also referred to as the
Fluctuation-Theorem \cite{GC95}. A key aspect of the FT theorem is
the occurrence of rare negative fluctuations in power. This has
been indirectly tested in fluid systems \cite{CL98,GGK01}, and
more recently direct measurements have been made in a shaken,
granular system \cite{FM03}. The second approach focuses on the
form of the probability distribution function, or distribution
function for short. One prediction is that the generalized
Fisher-Tippit-Gumbel (gFTG) distribution function describes power
fluctuations \cite{BCFH00}. Experiments using electroconvection
have demonstrated an interesting evolution of the distribution
function from Gaussian behavior to behavior that is consistent
with the gFTG distribution as the distance from equilibrium is
varied \cite{TG03}.

Another class of systems that have been used heavily for the study
of nonequilibrium fluctuations are soft-matter systems. A striking
observation that arises from these studies is the similarity
between the observations of fluctuations in foam
\cite{GD95,DK97,KE99,LTD02,PD03}, granular materials
\cite{MHB96,HBV99prl,HB03,FM03}, and colloidal suspensions
\cite{LDH03}. In the examples cited here, the fluctuations in
stress under conditions of steady applied strain rate were
studied. Similarities in various measures of the fluctuations
exist despite qualitative differences in the ``microscopic''
structure of the materials. For example, foam consists of gas
bubbles with liquid walls. For flowing foam, viscous dissipation
in the fluid walls is the major source of energy loss. For
comparison, the dissipation in granular systems is usually due to
dry friction between particles. Another potentially significant
difference is the the degree to which the particles or bubbles are
packed. Because of the solid nature of granular particles, they
are generally studied at packing densities near or below random
close packed. In contrast, foam ``melts'' at packing densities
near random close packed, and studies focus on packing densities
greater than the melting transition. Given these ``microscopic''
differences, it is worth commenting on some of the similarities
that are observed in previous studies.

For foam, there is both a rich history of theoretical modelling
\cite{KNN89,KOKN95,OK95,WBHA92,HWB95,D95,D97,JSSAG99,TSDKLL99} and
experimental studies \cite{GD95,DK97,LTD02,PD03,KE99} that deals
with the issue of fluctuations. The focus is on predicting the
probability distribution of energy (or stress) drops, referred to
as energy (stress) drop distributions. (Typically, simulations
focus on energy drops because it is relatively easy to compute the
total energy of the system. In contrast, experiments tend to focus
on measurements of stress, as this is the more accessible
quantity.) The focus on the energy (or stress) drop distributions
was originally motivated by potential connections to
self-organized criticality and the observation of power-law
behavior \cite{KNN89}. In this context, the models of foam can be
divided into two main classes: models that predict power-law
distributions for stress or energy drops
\cite{KNN89,KOKN95,OK95,WBHA92,HWB95} and models that predict a
definite average stress drop size \cite{D95,D97,JSSAG99,TSDKLL99}.
Experimental evidence from model two-dimensional systems generally
supports a distribution with a well-defined average
\cite{GD95,DK97,LTD02,PD03}. However, there is indirect evidence
for power-law distributions under conditions of quasi-static shear
\cite{KE99}. One issue that has not been thoroughly explored in
experiments is the possibility of the development of power-law
behavior near the melting transition. There is some evidence in
simulations exhibiting a well defined average energy drop that the
average energy drop increases as one approaches the melting
transition. This might indicate that the distribution is
approaching a power law \cite{TSDKLL99}.


For granular materials, there have been limited measurements of
stress drop distributions. Behavior similar to that observed in
foam is reported for a system of elastic disks that are slowly
sheared \cite{HB03}. In general, the distribution is consistent
with the existence of a characteristic size for the stress drops.
The system reported on in Ref.~\cite{HB03} exhibits a continuous
transition at which stress chains first extend across the system
\cite{HBV99prl}. Near this transition point, there is evidence of
possible power law behavior for the distribution of stress drops.

More commonly, the focus in granular systems has been on the
probability distribution for the stress itself
\cite{MHB96,HBV99prl,FM03}, as was also studied in colloidal
systems \cite{LDH03}. (For the purposes of this paper, this will
be referred to as the stress distribution, in contrast to the
stress drop distribution.) For all of these systems, one
surprising feature is how large the stress variations can be
\cite{MHB96,LDH03}. In one case, stress was measured normal to the
flow in a localized region of the system. For different particle
sizes, small variations in a general Maxwellian stress
distribution were observed, but all of the distributions had
essentially exponential tails for stress greater than the average
stress \cite{MHB96}. The nature of the stress distributions
appeared to depend strongly on the nature of the stress chains in
the granular material \cite{MHB96}. Another case measured stress
distributions in a colloidal system as a function of strain rates.
For low strain rates, gaussian distributions in stress were
observed. For higher strain rates extreme value distribution
statistics were measured with a well defined maximum value
\cite{LDH03}.


As a bridge between foam and granular systems, we have carried out
studies of fluctuations using a modification of the bubble raft
\cite{AK79}: a bead raft (plastic beads floating on the surface of
a fluid substrate). The plastic beads provide a connection with
granular systems. However, their individual density is such that
they are mostly submerged below the fluid substrate. Therefore,
there is a fluid layer between the beads, analogous in some ways
to the fluid walls in a foam. In this regard, the system is
closest to a two-dimensional colloidal system. However, in
contrast to typical colloidal systems used to study flow (see for
example, Ref.~\cite{LDH03}) the particles used here are much
larger (mm in size), so thermal fluctuations play no role. Also,
because the particles are only partially submerged, the role of
the lubrication of the fluid subphase is not well understood yet.

The bead raft is essentially a two dimensional system with free
boundary conditions perpendicular to the surface. The system has
an easily adjustable dispersity, packing fraction, strain rate,
and interaction potential.  We report on measurements of stress
during steady shear including average stress values (which
determine the flow regime for a given strain rate), stress
distributions, and stress drop distributions. By measuring both
the stress and the stress drop distributions, we are able to make
comparisons with the stress drop distributions observed in foam
and the stress distributions studies in granular matter. Using the
stress distributions, we can make indirect comparison with the
power dissipated distributions observed in turbulent flows.

\section{Experimental Details}

In our experiments, we use a centrosymmetric trough that can
measure stress with a freely hanging torsion pendulum. This
apparatus is described in detail in Ref.~\cite{app}. It consists
of two concentric cylinders oriented vertically.  The outer
cylinder consists of 12 individual ``fingers" that can be expanded
and compressed between 6 and 12 cm to adjust the packing fraction
of the beads. This barrier can be rotated in either direction at a
constant angular speed in the range of 0.0005 rad/sec to  0.4
rad/sec. The inner cylinder is a Teflon rotor that ``grips" the
material by the insertion of eight 5 mm aluminum ``paddles".  The
rotor has a diameter of 2.7 cm and is suspended by a torsion wire
with a torsion constant, $\kappa = 3020\ {\rm dyne\ cm/rad}$.
Induced voltage in a coil attached to the pendulum was used to
determine the angle of the pendulum.  When the outer barrier is
rotated we can measure the torque on the inner cylinder and the
corresponding stress on the inner cylinder. For the rest of the
paper, unless otherwise indicated, the stress refers to the stress
on the inner cylinder.

The system consists of spherical plastic beads localized to the
surface of a fluid by their buoyancy.  Their density is $0.95\
{\rm g/cm^3}$, allowing each sphere to float at the fluid's
surface. Unless noted, a bidisperse mixture of beads was used: 300
0.25 inch beads and 300 0.1875 inch beads. The simplest definition
of packing fraction was used: the number of beads of each size
times the maximal cross sectional area divided by the area of the
system. At the highest compressions, this does not account for
possible displacements perpendicular to the fluid surface. The
system was thoroughly mixed and a disordered two dimensional
system was created in the apparatus between the barriers. Because
some signs of aging on short time scales were found, all
measurements were performed after 1 hour of rotation in the same
direction. Over the time scale of our experiments, we found no
signs of coarsening or other settling phenomena common with some
agitated polydisperse granular systems.

The substrates used ranged from pure water to pure glycerine, and
included glycerine/water mixtures. The variation in substrate
effectively changes the interaction between the beads because of
changes in the properties of the fluid layer between particles.
For all cases, coupling of the beads to the substrate was
negligible. This was tested by removing the outer layer of beads
from contact with the outer barrier. In this case, rotation of the
outer barrier yielded zero stress on the inner rotor, and no
motion of the beads was detected in this case.

For many of the experiments, the system was video taped to monitor
the general behavior of the beads. Except for the highest packing
fractions (as discussed below), there was no discernable motion of
the beads out of the plane of the fluid surface. However, this
does not completely rule out that some of the release of stress
was due to slight motions of the beads in this direction. Such
releases in stress may be important if quantitative comparison
with purely two-dimensional models are carried out in the future.
However, for the main results of this paper, which is focused on
the degree of {\it universality} of the fluctuations, the details
of the source of stress release are not relevant.

\section{Results}

\begin{figure} [htb]
\includegraphics[width=8cm,keepaspectratio=true]{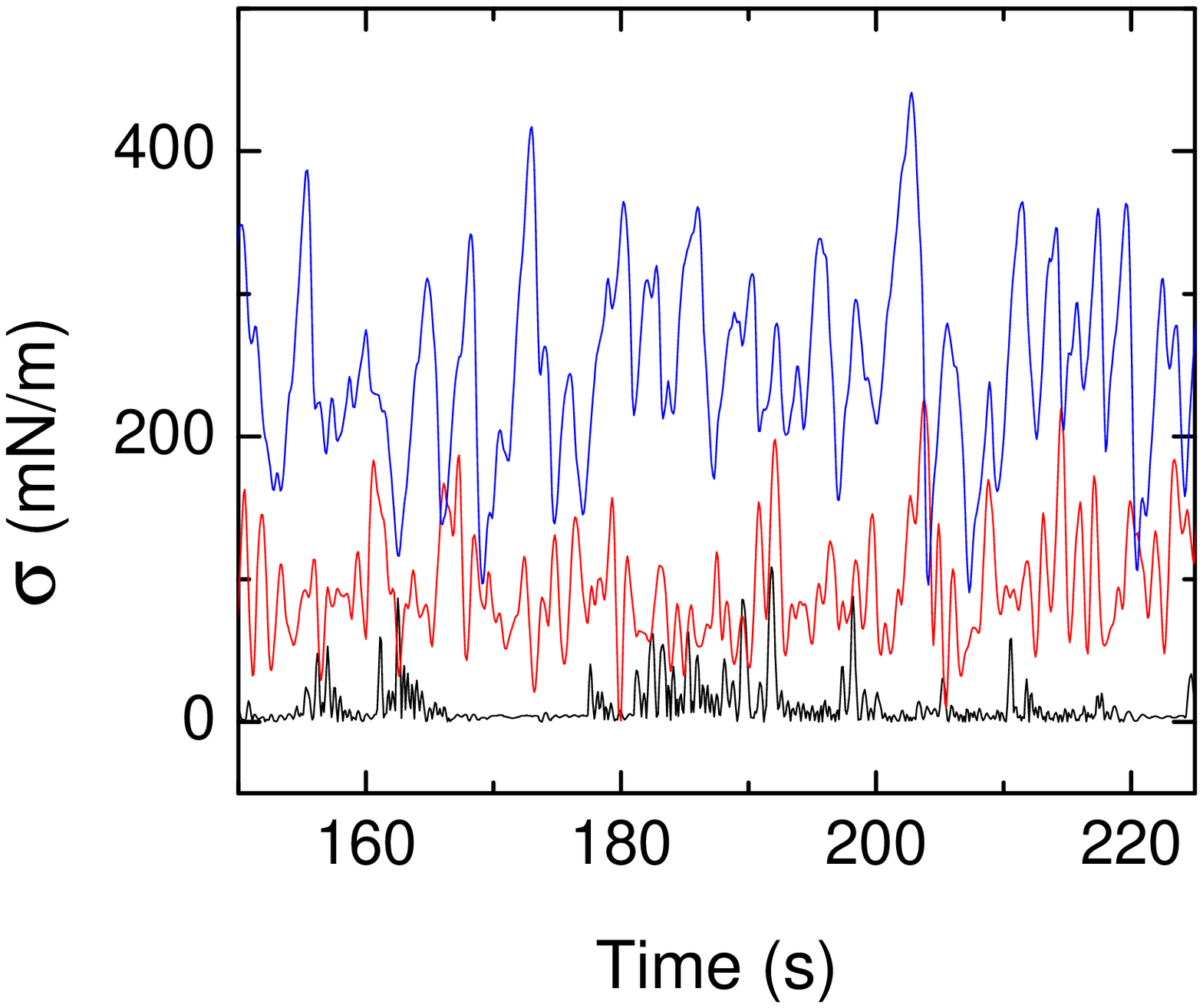}
\caption{\label{stresstime} A plot of stress on the inner cylinder
versus time during steady rotation of the outer barrier. The
characteristic stress relaxations are evident. The packing
fractions for increasing average stress are 0.705 (black), 0.74
(red) and 0.78 (blue).}
\end{figure}

Figure~\ref{stresstime} shows typical time series of the stress on
the inner rotor as the outer barrier is rotated for three
different packing fractions. The packing fractions were selected
to illustrate three qualitatively different behaviors. For the
lowest packing fraction (solid black line, lowest curve), the
stress is predominately zero. This is due to the fact that under
constant shear, the particles are driven toward the outer
boundary. For sufficiently low densities, this results in exactly
zero stress on the inner boundary, as no particles are in contact
with it. As the packing fraction is increased, there is a density
regime for which particles intermittently contact the inner rotor.
This results in the occasional spikes in the stress.

The second behavior is illustrated by the red line (middle curve).
This occurs when there are always particles in contact with the
inner rotor, but the fluctuations occasionally result in the
stress dropping back to zero. Finally, at high enough packing
fraction (blue curve, highest average stress), the stress is
always nonzero during the time for which the flow is observed. The
rest of the results focus on the average stress value, the
probability distribution of the stress, and the stress drop
distributions, as calculated from time series similar to those
illustrated in Fig.~\ref{stresstime}.

\begin{figure} [htb]
\includegraphics[width=8cm,keepaspectratio=true]{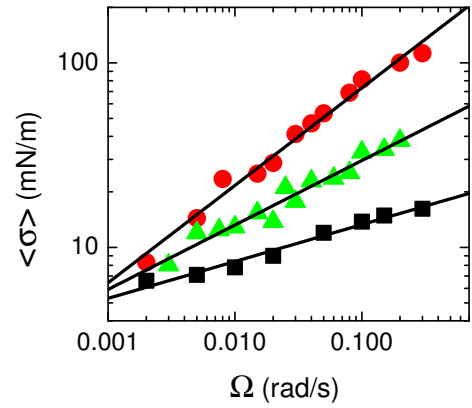}
\caption{\label{stressvrot} A plot showing average stress values
versus rotation rate for three different substrates: pure
glycerine ($\bullet$); 70 \% glycerine/ 30 \% water mixture
($\blacktriangle$); and pure water ($\blacksquare$). The solid
lines indicate the consistency between the data and power law
behavior.}
\end{figure}

The first question that is addressed is the dependence of the
average stress on the rotation rate and the density.
Figure~\ref{stressvrot} presents the results for the average
stress versus rotation rate for three different substrates. For
all three cases, the behavior is consistent with that of a power
law fluid. A power law fluid is one in which the viscosity,
$\eta$, as a function of strain rate, $\dot{\gamma}$ is given by:
$\eta \propto \dot{\gamma}^{(m-1)}$. For the Couette geometry,
$\dot{\gamma} \propto \Omega$, where $\Omega$ is the outer
rotation rate. This gives $<\sigma> = \eta \dot{\gamma} \propto
\Omega^m$. From this, we find that for glycerine $m = 0.53 \pm
0.02$. For  $70\%$ glycerine $30\%$ water mixture, $m = 0.35 \pm
0.02$, and for pure water, $m = 0.20 \pm 0.01$.

\begin{figure} [htb]
\includegraphics[width=8cm,keepaspectratio=true]{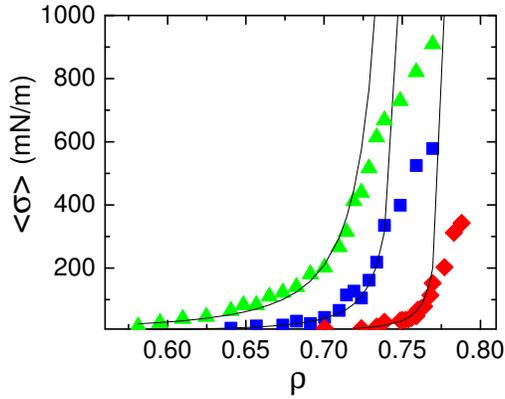}
\caption{\label{stressvden} A plot of the average stress value
versus packing fraction for three different situations: rotation
rate of 0.1 rad/s with pure glycerine substrate
($\blacktriangle$); rotation rate of 0.02 rad/s with pure
glycerine substrate ($\blacksquare$); and rotation rate of 0.1
rad/s with pure water substrate ($\blacklozenge$). The solid lines
are a fits to Eq.~\ref{divergence}, and they illustrate the
diverging nature of the stress as the material is compressed.}
\end{figure}

Before discussing these results further, it is useful to consider
the average stress as a function of packing fraction, as
illustrated in Fig.~\ref{stressvden}.  For this system, there are
a number of transitions as a function of packing fraction. As
already discussed with regard to Fig.~\ref{stresstime}, for small
enough packing fractions there is no transfer of stress between
the inner and the outer barriers. The transition from this regime
to non-zero average stress is somewhat poorly defined due to
long-time transients. Figure~\ref{stressvden} only covers
densities for which the stress did not go to zero over
experimental time scales of a few hours. The second transition is
the ``buckling'' of the bead raft. This is indicated by the change
in slope at higher packing fraction and is confirmed by imaging
the bead raft.

Before the buckling transition occurs, the average stress as a
function of density is consistent with a diverging power law,
suggestive of an approach to the ``jamming'' transition
\cite{LN98}. Therefore, for the range of density of interest
(below buckling and with nonzero average stress), the system is
below the jamming transition and should be in a purely fluid
state. This is consistent with the results presented in
Fig.~\ref{stressvrot}, from which there is no evidence for a yield
stress. In contrast, the bubble raft is above the jamming
transition and clearly exhibits a yield stress \cite{LTD02,PD03}.

A critical packing fraction $\rho_{c}$ for the jamming transition
is determined by fitting the data to

\begin{equation}
\label{divergence}
\langle\sigma\rangle = \frac{\sigma_o}{(1 -
(\rho/\rho_c))^n}.
\end{equation}

The fits are given by the solid lines in Fig.~\ref{stressvden}. It
should be noticed that this fit focuses on the divergence of the
stress at higher packing fraction. This is opposite of the
definition of the critical packing fraction reported in
Ref.~\cite{HBV99prl}, where the interest was the initial
transition to nonzero stress.

Figures~\ref{stressvrot} and \ref{stressvden} reveal a dependence
of two characterizations of the average properties of the system
on the choice of substrate: the exponent of the power-law
viscosity ($m$) and the critical density for jamming ($\rho_{c}$).
All of the data can be fit with the same exponent of the stress
divergence ($n = 2$). The general trend is for $m$ to increase as
the concentration of glycerine is increased, and for $\rho_c$ to
decrease as parameters are decreased. The values of $m$ were
already reported. For $\rho_c$, some examples are $\rho_c = 0.74$
and $\rho_c = 0.78$ for the pure glycerine and water substrates,
respectively. An interesting feature of $\rho_c$ is that it
appears to be independent of rotation rate, even for the glycerine
substrate. The dependence of the parameters on substrate
presumably reflects changes in the interactions between the beads
due to the changing properties of the substrate. This is possible
because the substrate coats a significant fraction of the beads
and forms a thin film between particles. Therefore, it is
reasonable to consider variations in substrate composition as
variations in particle interactions.

\begin{figure} [htb]
\includegraphics[width=8cm,keepaspectratio=true]{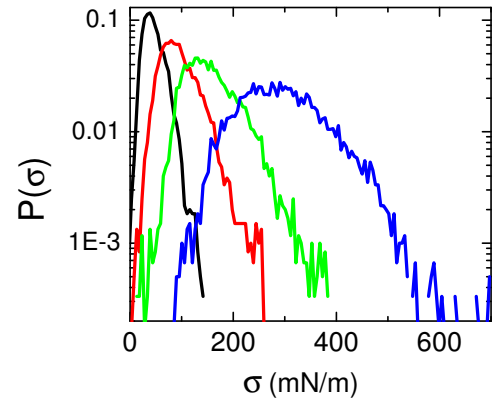}
\caption{\label{stressdist} A plot of the probability of a
particular stress value for beads on pure water with a rotation
rate of 0.1 rad/s. The four curves correspond to packing fractions
of (in order of increasing mean): 0.71 (black); 0.74 (red); 0.75
(green); and 0.78 (blue).}
\end{figure}

\begin{figure} [htb]
\includegraphics[width=8cm,keepaspectratio=true]{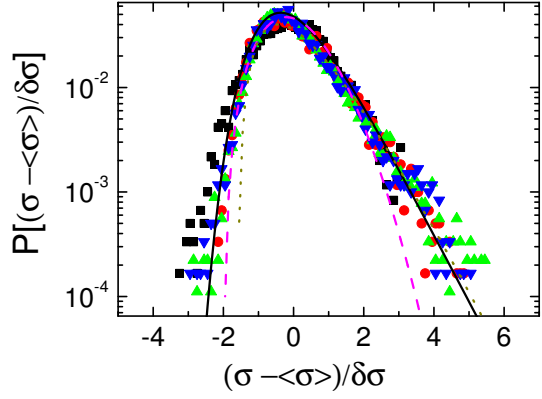}
\caption{ \label{normstress} The probability of $(\sigma -
<\sigma>)/\delta\sigma$ is plotted, where $\delta\sigma$ is the
standard deviation of the stress distribution. This is referred to
as the normalized stress distribution. The different substrates
and packing fractions are as follows: water substrate and packing
fraction 0.71 (blue $\blacktriangledown$); water substrate and
packing fraction 0.77 (green $\blacktriangle$); packing fraction
0.68 and glycerine substrate (red $\bullet$); and a monodisperse
system on water susbtrate (black $\blacksquare$). The lines are
fits to three different possible distributions, as described in
the text: the gFTG (solid line), two-dimensional Maxwell Boltzmann
(dashed line), and one-dimensional Maxwell Boltzmann (dotted
line)}
\end{figure}

Figure~\ref{stressdist} displays the probability distribution for
the stress for four different packing fractions on a water
substrate. As the packing fraction increases, the average stress
($<\sigma>$) and the variance in stress ($\delta \sigma$)
increase. Figure~\ref{normstress} shows normalized probability
distributions, where we have plotted the probability of $(\sigma -
<\sigma>)/\delta \sigma$ for a number of different situations.
There are two striking features of the normalized distributions.
First, for all of the parameter values for which we made
measurements, the normalized distributions essentially collapse on
a single curve. Second, the curve is asymmetric, with an
exponential tail for stresses greater than the average stress.

In order to make comparisons with other work on stress
distributions, we attempted to fit the data to three different
models. The first model is based on the gFTG distribution function
used to describes power fluctuations \cite{BCFH00}:
\begin{equation}
\Pi(\sigma)=K(e^{x-e^{x}})^{a},
\end{equation}
where $x=b(\sigma-s)$ and $a=\pi/2$.

The parameters in this function are set by the criteria of unit
area, zero mean, and unit variance. The plot in
Fig.~\ref{normstress} uses values for the parameters that are
expected based on Ref.~\cite{BCFH00}: $s = 0.37$ and $b = -0.94$.
The only difference is that the constant b is negative because our
distribution is flipped with regard to the asymmetry measured in
turbulent systems.

We also fit our plot to distributions of the form
\begin{equation}
P(\sigma)=(\sigma-\sigma^*)^{2}e^{-(\sigma-\sigma^*)^{n}/
<\sigma>}
\end{equation}
for n = 1 and 2. These are chosen to compare with the results from
Ref.~\cite{MHB96} and a Maxwell-Boltzmann distribution,
respectively. Here, $\sigma^*$ was adjusted to get the best fit to
the data. Not surprisingly, the gFTG distribution and distribution
used in Ref.~\cite{MHB96} best capture the exponential tail for
high stress and provide the closest fit to the data.

\begin{figure} [htb]
\includegraphics[width=8cm,keepaspectratio=true]{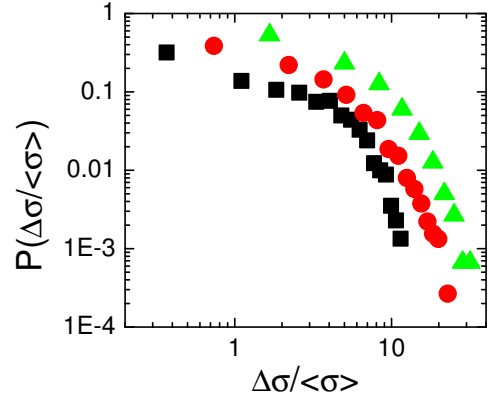}
\caption{\label{dropdist} A plot of the probability distribution
for stress drops normalized by the average stress ($\Delta
\sigma/<\sigma>$ for three different densities: $\rho = 0.735$
($\blacktriangle$); $\rho = 0.75$ ($\circ$); and $\rho = 0.765$
($\blacksquare$).}
\end{figure}

\begin{figure} [htb]
\includegraphics[width=8cm,keepaspectratio=true]{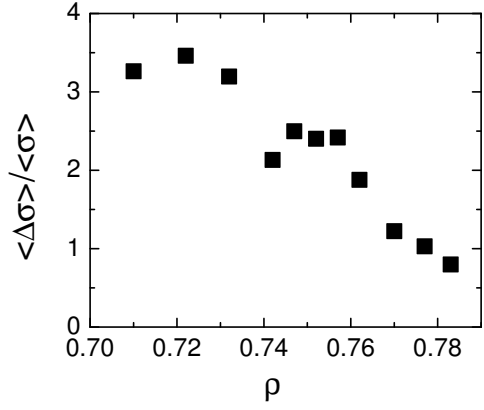}
\caption{ \label{avedrop} A plot of the average stress drop,
normalized by the average stress, ($<\Delta \sigma>/<\sigma>$
versus packing fraction. This is for the bidisperse system on
water substrate.}
\end{figure}

Figure ~\ref{dropdist} shows the distribution of stress drops for
three packing fractions. The general trend is similar to that
observed for foam. There is an exponential cutoff in the stress
drop distribution that results in a well-defined average stress
drop. Unlike the normalized stress distribution, the average
stress drop increases as one approaches the jamming transition
from below. However, the average stress drop does not diverge as
quickly as the average stress itself. Therefore, the average
stress drop normalized by the average stress actually decreases as
a function of density (see Fig.~\ref{avedrop}).

\section{Discussion}

We have presented results for the response of a plastic bead raft
floating on a fluid substrate to the application of constant rate
of strain. The system was selected to provide a bridge between
similar studies in foam and granular materials by combining
features of both materials, as more traditional collodial systems
do. The system was studied in a Couette geometry in which the
outer cylinder was rotated at a constant angular speed and the
stress on the inner cylinder was measured. A number of properties
were measured as a function of packing fraction (or density),
applied strain rate (or rotation rate), and interaction strength
(or substrate composition). The main focus of the study was the
behavior of the fluctuations in stress. In this regard, two
quantities were characterized: the distribution of stress and the
distribution of stress drops.

The results for the stress drop distributions are very similar to
those found in foam \cite{LTD02,PD03} and granular material
\cite{HB03}. There is evidence for power-law behavior for small
stress drops, but there is an exponential cutoff. The net result
is a well-defined average stress drop. The average stress drop
does increase as the packing fraction is increased, but not as
fast as the average stress. There was no strong evidence for an
approach to a power-law behavior at either the high or low
critical densities. However, this might have been due to
difficulties with approaching close enough to these transitions.
For low packing fractions, the system became highly inhomogeneous,
with beads concentrating toward the outer cylinder. At the high
end, there was the cross-over to stress releases due to bead
motion into the third dimension. Future work will be needed to
explore the detailed behavior near these transitions.

With regard to the distribution of stress itself, some interesting
``universal'' features are observed. The most obvious feature is
the exponential tail in the stress distribution that is consistent
with observations in both granular systems \cite{MHB96} and highly
driven fluids \cite{TG03}. As with the driven fluid system, it is
intriguing to note that the gFTG distribution function, proposed
in Ref.~\cite{BCFH00}, is consistent with the data. The one
difference is that the exponential tail in our data is for stress
greater than the average stress. This is similar to the granular
systems, in which stress is also the measured quantity
\cite{MHB96}. This difference can be understood in terms of the
connection between stress transmitted to the inner rotor and the
power {\it dissipated} by the system. Because a constant power is
input from the outer rotor, when the stress on the inner cylinder
is higher than average, the power dissipated by the system is
lower than average. Additional theoretical work is needed to fully
understand why, and under what conditions, the gFTG distribution
applies to this system. Also, it should be noted that there is
still sufficient scatter in the data that it is not possible to
rule out other distributions with exponential tails, such as the
function used in Ref.~\cite{MHB96}.

One feature of the stress distributions that we were not able to
access with our system was the extremely high strain rate limit.
For colloids in this limit \cite{LDH03}, the stress distribution
developed an interesting structure that included a plateau. This
behavior was never observed in our system. Future work will
attempt to reach the necessary high strain rates, as the beads at
the air-water interface is essentially a colloidal system.

Finally, the average stress was found to diverge as one approached
a critical density as the density was increased at constant strain
rate. This represents a diverging viscosity for the system. One
way to understand this behavior is within the context of the
proposed ``jamming'' transition \cite{LN98,TPCSW01,ITPJamming}.
Jamming is the crowding of constituent particles, disabling their
kinetics and further exploration of phase space
\cite{LN98,TPCSW01,ITPJamming}. When a system develops a yield
stress or extremely long relaxation time, it jams \cite{COH02}.
The jamming phase diagram proposes that one can approach the
jammed state by either varying density, temperature, or shear
stress \cite{LN98}. An interesting feature of the jamming
transition is that for finite size systems in the absence of
shear, there is a distribution of critical densities that depend
on the particle configurations \cite{COH01,COH02,CSLN03}. An open
question is how this distribution of jamming transitions impacts
the behavior of the system under steady shear. It is easy to
imagine that the shear causes the system to explore configuration
space, randomly moving between jammed and unjammed states. This
exploration of phase space would be a source of stress
fluctuations. A future challenge is to understand the connection
between the average stress, the fluctuations, and the distribution
of jammed states. Future experiments exploring the impact of
system size will be critical to answering this intriguing
question.

\begin{acknowledgement}

This work was supported by Department of Energy grant
DE-FG02-03ED46071, the Research Corporation and Alfred P. Sloan
Foundation. The authors thank Robert Behringer, Corey O'Hern, and
Andrea Liu for useful discussions.

\end{acknowledgement}


\clearpage

\end{document}